\newcounter{myctr}
\def\myitem{\refstepcounter{myctr}\bibfont\noindent\ifnum\themyctr>9\else\phantom{0}\fi\hangindent17pt\themyctr.\enskip}

\documentclass[12pt]{iopart}
\usepackage{iopams}
\usepackage{amsfonts}
\usepackage{graphicx}
\usepackage{epsfig}
\usepackage{graphics}

\newcommand{\prl}[3]{Phys.\ Rev.\ Lett.  \textbf{#1}, #2 (#3)}

\newcommand{\bra}[1]{\bigl\langle #1 \bigr|}
\newcommand{\ket}[1]{\bigl| #1 \bigr\rangle}

\makeatletter
\begin{document}



\title{New aspects of the purity and information of an entangled qubit
pair.\\
  }

\author{Nasser Metwally  }

\address{Math. Dept. , College of Science,Bahrain University,
 32038 Kingdom
of Bahrain.\\
Mathematics Department, Faculty of Science, South Valley
University, Aswan, Egypt.\\
Nmetwally@gmail.com}


\begin{abstract}
In this article, we investigate the dynamics of the purity of the
entangled of 2 two-level atoms interacting with a single quantized
electromagnetic field. We show that how the purity of the qubit
pairs depends on initial state of the atomic system. It is found
that the superposition case is the best choice to generate
entangled states with high purity and hence high entanglement. It
is clear that the purity of one qubit can purified by the expense
of the other pair through the phenomena of purity swapping. The
mean photon number plays an important role in increasing the
purity.  The robustness of the quantum channel is investigate in
the presences of individual attacks, where we study the
separability of these channels and evaluate its fidelity. Finally,
we use these channels to perform the original coding protocol by
using theses partial entangled states. We find that Bob can gets
the codded information with reasonable percentage. The inequality
of security is tested, where we determine the interval of times in
which Alice and Bob can communicate secure. These intervals depend
on the type of error and the structure of initial atomic system.
\end{abstract}
\textbf{keywords:}Qubit, Entanglement, Purity,  Dense coding.
\maketitle

\section{Introduction}
Evolution of atoms interacting with a coherent field is an
interesting topic in quantum optics~\cite{phoenix}. The Jaynes-
Cummings model, JCM, is the simplest model describes the
interaction of a single two -level atom with a single quantized
electromagnetic field and can be realized experimentally
\cite{Brune}. Nowadays, related studies are concerned with quantum
information \cite{You,Metwally} and computations \cite{Bige}.
 In this work, we consider simple model of a two two-level atoms
interacting with a single mode radiation field. Due to the
interaction, the behavior of the two atoms swap from separable to
entangled and vis versa. In our calculations, we consider only the
entangled case of the two qubit state. On the other hand, since
most of quantum information processing requires entangled state
with high degree of purity and hence a degree of entanglement, we
focus here on the effect of the structure of the initial atomic
system on the degree of purity i.e, we answer the questions, what
is the best choice of the initial atomic state to generate an
entangled state with high purity? and how the purities of the
entangled qubit pairs and their individual subsystems evolves with
time?. Since one can use these states as channels to code
information,  we investigate the robustness of theses channel in
the presences of the individual attacks. Finally, we use theses
quantum channel to perform original coding protocol
\cite{Bennett}. Also,  the possibility of  sending  a secure
information between two users is investigated, where we determine
the secure interval of times, in which the users can communicate
secure.

The article is arranged as follows: In sec.$2$, we describe the
model and its time evolution. Also we introduce the final state of
the density operator in the computational basis. The behavior of
the purity is the subject of subsection $3$, where the
calculations  are performed only when the density operator behaves
as an entangled state. Sec.$4$ is devoted to investigate the
robustness of the quantum channel is in the presences of
eavesdropper, where we consider Eve applies the individual attacks
strategy. In sec. $5.1$, we calculate the average amount of
information gained from the coded information. Also the
possibility of sending a secure information is investigate in
sec.$5.2$. Finally we discuss our result in sec.$6$.

\section{The system and its evolution}
The Hamiltonian which describes a system of a two two -level atoms
each consisting of states $\ket{e}$ and $\ket{g}$ coupled to a
single mode radiation field in the rotating wave approximation is
given by
\begin{equation}
H=\omega(a^\dagger  a+\sum_{i=1}^{2}\sigma_z^{i})+ \sum_{i=1}^{2}[
\lambda_i(a^\dagger\sigma_{-}^{(i)}+\sigma_{+}^{(i)}a)]
\end{equation}
where $a(a^{\dagger})$is the annihilation (creation) operator of
the field mode, $\sigma_{\pm}^{i}$ and $\sigma_z^{i}$ the
parameters $\lambda_i$ are the atom-field coupling constant
$\omega$, is the atomic transitions  and the field mode frequency.
The first term in Eq.$(1)$ represents the free-Field and the
non-interacting atoms, while the second term stands for the
interaction Hamiltonian, $H_{int}$. This model has been solved
analytically for some special cases and for a general case
\cite{Dung}. In this work, we introduce a direct solution for this
model. For simplicity, we consider the case of identical atoms
i.e.$\lambda_1=\lambda_2$. Assume that the cavity field is
initially prepared in a coherent state
$\ket{\psi(0)}_{f}=\sum_{n=0}^{\infty}{q_n\ket{n}}$, where
$q_n=exp(-\frac{\bar n}{2}) \frac{\bar
n^{\frac{n}{2}}}{\sqrt{n!}}$. For the atomic system, we consider
the first atom is in its excited state i.e.
$\ket{\psi(0)}_1=\ket{e}_1$ and the other atom is in a
superposition state $\ket{\psi(0)}_2=a\ket{e}_2+b\ket{g}_2$, where
$1$ stands for the first atom, 2 for the second atom  and
$|a|^2+|b|^2=1$. So, we can write the initial state of the two
atoms as
$\ket{\psi(0)}_{12}=\ket{e}_1\otimes(a\ket{e}_2+b\ket{g}_2)$ and
consequently the initial state of the combined system is given by,
\begin{equation}
\ket{\psi_0}=\sum_{n=0}^{\infty}q_n\Bigr\{\ket{n}\otimes(a\ket{e}_1\otimes\ket{
e}_2+b\ket{e}_1\otimes\ket{g}_2)\Bigl\}.
\end{equation}
At any time $t > 0$, the atom-field state is described by the
state,
\begin{equation}
\ket{\psi(t)}=e^{-iHt}\ket{\psi(0)}.
\end{equation}
 Since we are interested in the behavior
of the purity and information contained in the entangled two
atoms, we trace over the field state. After some straightforward
calculations one can get the density operator of the two atoms
explicitly in the computational basis as
$\ket{00},~\ket{10},~\ket{01}$ and $\ket{11}$:
\begin{eqnarray}
\rho(t)_{12}&=&|c_n^{(1)}|^2\ket{00}\bra{00}+c_n^{*(1)}c_{n+2}^{(2)}\ket{10}\bra{00}
+c_n^{*(1)}c_{n+1}^{(3)}\ket{01}\bra{00}+c_n^{*(1)}c_{n+2}^{(4)}\ket{11}\bra{00}
\nonumber\\
&+&c_n^{(1)}c_{n+2}^{*(2)}\ket{00}\bra{10}+|c_n^{(2)}|^2\ket{10}\bra{10}
+c_n^{*(2)}c_{n}^{(3)}\ket{01}\bra{10}+c_n^{*(2)}c_{n+1}^{(4)}\ket{11}\bra{10}
\nonumber\\
&+&c_n^{(1)}c_{n+1}^{*(3)}\ket{00}\bra{01}+c_n^{(2)}c_{n}^{*(3)}\ket{10}\bra{01}+
 |c_n^{(3)}|^2\ket{01}\bra{01}+c_n^{*(3)}c_{n+1}^{(4)}\ket{11}\bra{01}
\nonumber\\
&+&c_n^{(1)}c_{n+2}^{*(4)}\ket{00}\bra{11}+
c_n^{*(2)}c_{n+1}^{(4)}\ket{10}\bra{11}+
 c_n^{(3)}c_{n+1}^{*(4)}\ket{01}\bra{11}+
  |c_n^{(4)}|^2\ket{11}\bra{11},
\nonumber\\
\end{eqnarray}
where,
\begin{eqnarray}\label{Deeg}
c_{n}^{(1)}(t)&=&-\sum_{n=0}^{\infty}q^2_n\frac{\gamma_n}{\sqrt{\mu_n}}
\left(\frac{b\beta_n}{\sqrt{\mu_n}}(1-\cos t\sqrt{\mu_n}) +ia\sin
t\sqrt{\mu_n}\right),
\nonumber\\
c_{n}^{(2)}(t)&=&\sum_{n=0}^{\infty}q^2_n\left(
\frac{2a}{\mu_n}(\beta^2_n+\gamma^2_n)\sin^2t\sqrt{\mu_n}+a\cos
t\sqrt{\mu_n}-i\frac{2\beta_n}{\sqrt{\mu_n}}\sin
t\sqrt{\mu_n}\right),
\nonumber\\
c_{n}^{(3)}(t)&=&-\sum_{n=0}^{\infty}q^2_n\frac{\sin
t\sqrt{\mu_n}}{\sqrt{\mu_n}}\left(\frac{2a}{\sqrt{\mu_n}}(\beta^2_n+\gamma^2_n)
+b\beta_n)\right),
\nonumber\\
c_{n}^{(4)}(t)&=&\sum_{n=0}^{\infty}q^2_n
 \frac{\cos
t\sqrt{\mu_n}}{\sqrt{\mu_n}}\left(\frac{2b\gamma^2_n}{\sqrt{\mu_n}}(\mu_n-2\gamma^2_n)
-ia\beta_n\right),
\end{eqnarray}
with
\begin{equation}
\gamma_n=\sqrt{n+1},\quad \beta_n=\sqrt{n+2},\quad
\mu_n=2(\gamma^2_n+\beta^2_n), \quad\tau=\lambda t.
\end{equation}

\section{Purity}
Most of quantum information tasks require entangled pure state to
be  performed. This aim is a difficult to achieve in reality due
to the dechorence. In this section, we try to answer the following
question. How close the generated entangled state of the two atoms
to purity?. On the other hand, since the interaction depends on
the initial state of the two atoms, we study how the states of the
atoms behave individually. In our study, we consider the impurity
of the two atoms $\rho_{12},~\rho_1$ for the first atom and
$\rho_2$ for the second atom. Let us define the degree of impurity
as
\begin{equation}\label{purity}
\eta_i=1-tr{\rho_i^2},
\end{equation}
where $i=1,2,12$ see for example \cite{jex}. For the  density
operator $\rho_{12}$ the degree of impurity is given by,
\begin{eqnarray}
\eta_{12}&=&1-\biggl\{
|c_{n}^{(1)}|^2\left[|c_{n}^{(1)}|^2+2(|c_{n+1}^{(2)}|^2+|c_{n+1}^{(3)}|^2+
|c_{n+2}^{(4)})|^2)\right]
\nonumber\\
&+&|c_{n}^{(2)}|^2\left[|c_{n}^{(2)}|^2+2(|c_{n}^{(3)}|^2+|c_{n+1}^{(4)}|^2)\right]
\nonumber\\
&+&|c_{n}^{(3)}|^2\left[|c_{n}^{(3)}|^2+2(|c_{n+1}^{(4)}|^2)\right]+|c_{n}^{(4)}|^4\biggr\},
\end{eqnarray}
for the first atom is,
\begin{figure}[b]
  \begin{center}
\includegraphics[width=25pc,height=10pc]{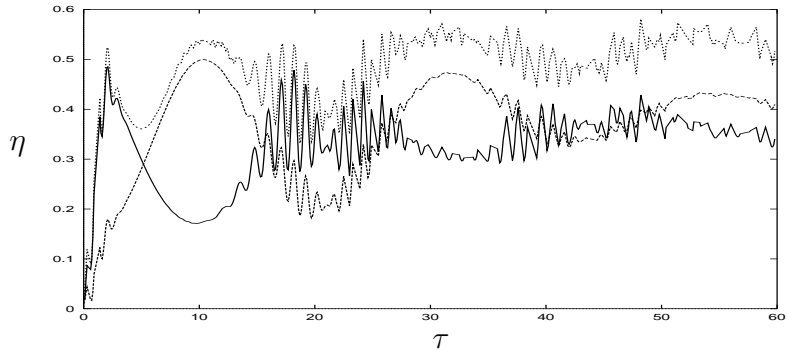}
  \put(-140,-10){$\tau$}
\put(-300,65){$\eta$}
  \caption{The degree of impurity as a function of the scaled time
  $\tau$ with $=\bar n=10$. The dot line for $\rho_{12}$ the solid line for $\rho_1$
   and the dash line for $\rho_2$. The atomic system is initially prepared
   in a superposition product state $a\ket{ee}+b\ket{eg}$, where
      $a =0.5$ and b=$\sqrt{1-a^2}$.}
  \end{center}
\end{figure}

\begin{eqnarray}
\eta_{1}&=&1-\biggl\{\left(|c_n^{(1)}|^2+|c_n^{(3)}|^2\right)^2
\nonumber\\
&+&2\biggl[|c_{n+1}^{(2)}|^2|c_n^{(1)}|^2
                 +|c_{n+1}^{(4)}|^2|c_n^{(3)}|^2
             |c_{n+1}^{(2)}|^2|c_n^{*(1)}|^2|c_n^{(3)}|^2|c_{n+1}^{(4)}|^2
\nonumber\\
&+&|c_{n+1}^{(4)}|^2 |c_n^{*(3)}|^2
|c_n^{(1)}|^2|c_{n+1}^{*(2)}|^2
             +\left(|c_n^{(2)}|^2+|c_n^{(4)}|^2\right)^2\biggr]\biggr\},
\end{eqnarray}
and for the second atom is,
\begin{eqnarray}
\eta_{2}&=&1-\biggl\{\left(|c_n^{(1)}|^2+|c_n^{(2)}|^2\right)^2
\nonumber\\
&+&2\biggl[|c_{n}^{(1)}|^2|c_{n+1}^{(3)}|^2
                 +|c_{n}^{(2)}|^2|c_{n+1}^{(4)}|^2
             |c_{n+1}^{(3)}|^2|c_n^{*(1)}|^2|c_n^{(2)}|^2|c_{n+1}^{*(4)}|^2
\nonumber\\
&+& |c_{n+1}^{(4)}|^2 |c_n^{*(2)}|^2
|c_n^{(1)}|^2|c_{n+1}^{*(3)}|^2
             +\left(|c_n^{(3)}|^2+|c_n^{(4)}|^2\right)^2\biggr]\biggr\},
\end{eqnarray}
where $c's$ are given by (\ref{Deeg}). Fig.$1$, shows the behavior
of the degree of impurity for three states, the first atom
$\rho_1$, the second atom $\rho_2$  and the state of the two atoms
$\rho_{12}$. In this investigation, we consider only the {\it
entangled} states of $\rho_{12}$ and its corresponding subsystems
$\rho_1$ and $\rho_2$. We consider that  the initial state of the
system is in a superposition state and the mean photon number
$\bar n=10$. From this figure, it is clear that when  the impurity
of the first atom is  maximum,  the impurity of the   second one
has a minimum impurity. This phenomena is called impurity {\it
swapping}. On other words, one can say that  the purity of one
qubit is increased at the expense of the other qubit. The behavior
of $\eta$ for $\rho_{12}$ and $\rho_2$ is the same, but the degree
of purity is different. For some intervals of time we obtain an
entangled stated of purity, $(1-\eta)>0.90$. This is an important
result for quantum information and computation, where these
applications need pure states with a high purity to be performed.
Also we can see that as the time evolves the degree of purity
decreases. This explain why the degree of entanglement decreases
as the time evolves.
\begin{figure}[t]
  \begin{center}
\includegraphics[width=25pc,height=11pc]{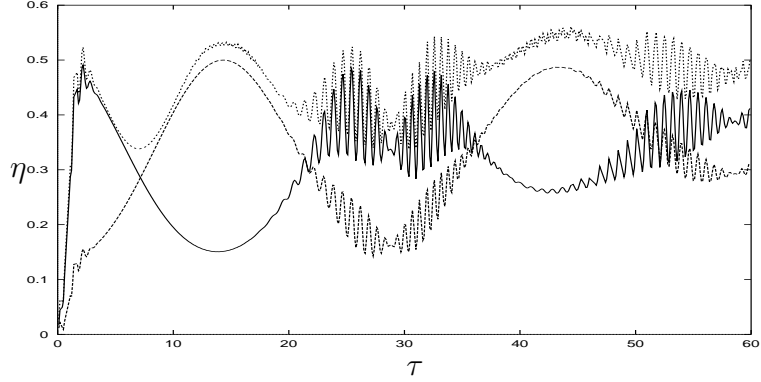}
  \put(-140,-10){$\tau$}
\put(-290,65){$\eta$}
  \caption{The same as Fig.$1$, but $\bar n=20$.}
  \end{center}
\end{figure}

Fig.$2$, shows the effect of the mean photon number on the degree
of purity. It is clear that as one increases $\bar n$. the degree
of purity increases. This is clear by comparing Fig.$1$ and
Fig.$2$, in addition to the usual behavior of Rabi oscillations is
seen. The only difference in these figures is that the number of
oscillations is different. This is due to the lost of phase to the
separable state, where in our calculation, we consider only the
{\it entangled states}. As $\bar n$ increases , a large modulation
in the oscillations and a clear shift from the usual Rabi
oscillations occurs. Also the degree of purity increases as the
values of $\bar n$ increases and the amplitudes of Rabi
oscillations decrease \cite{Aty}.
\begin{figure}
  \begin{center}
\includegraphics[width=25pc,height=10pc]{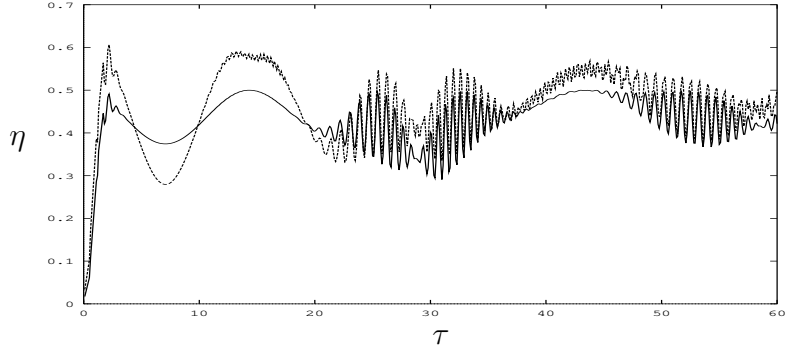}
  \put(-140,-10){$\tau$}
\put(-300,65){$\eta$}
  \caption{The same as Fig.$2$, but the atomic system is prepared initially in  exited state.}
  \end{center}
\end{figure}

In Fig.$3$,  we consider that the atomic system is initially
prepared  in excite state. In this case the degree of purity
decreases and hence the degree of entanglement. By comparing
Fig.$2$ and Fig.$3$, we can conclude that one can generate an
entangled state with high degree of purity and consequently high
degree of entanglement, if the atomic system is prepared initially
in a superposition product states. It is clear that the degree of
purity depends on the structure of the initial states of the two
atoms. From this, we can argue that, starting from a superposition
state as an initial state for the atomic system is recommended as
a best one of the nest choices.

\section{Robustness of the quantum channel } In this section, we
try to investigate the stability of the quantum channel in the
presence of the eavesdropper. Let us restrict ourself on the
individual attacks. For this strategy, Eve, has the ability to
access Alice's qubit and make a projective measurements along a
certain basis to get the encoded information. Also, we assume that
Eve sends another qubit to Bob by applying a unitary operators $
I,\sigma_x,\sigma_y,\sigma_z$ randomly on the original state. As
an example if Eve applies $\sigma_z$ on the travelling qubit of
Alice, then Bob will get a new state is defined by.
\begin{eqnarray}
\tilde\rho(t)_{12}&=&|c_n^{(1)}|^2\ket{00}\bra{00}-c_n^{*(1)}c_{n+2}^{(2)}\ket{01}\bra{00}
+c_n^{*(1)}c_{n+2}^{(3)}\ket{01}\bra{00}-c_n^{*(1)}c_{n+2}^{(4)}\ket{11}\bra{00}
\nonumber\\
&-&c_n^{(1)}c_{n+2}^{*(2)}\ket{00}\bra{10}+|c_n^{(2)}|^2\ket{10}\bra{10}
+c_n^{*(2)}c_{n}^{(3)}\ket{01}\bra{10}+c_n^{*(2)}c_{n+1}^{(4)}\ket{11}\bra{10}
\nonumber\\
&+&c_n^{(1)}c_{n+1}^{*(3)}\ket{00}\bra{01}+c_n^{(2)}c_{n}^{*(3)}\ket{10}\bra{01}+
 |c_n^{(3)}|^2\ket{01}\bra{01}-c_n^{*(3)}c_{n+1}^{(4)}\ket{11}\bra{01}
\nonumber\\
&-&c_n^{(1)}c_{n+2}^{*(4)}\ket{00}\bra{11}+
c_n^{*(2)}c_{n+1}^{(4)}\ket{10}\bra{11}-
 c_n^{(3)}c_{n+1}^{*(4)}\ket{01}\bra{11}+
  |c_n^{(4)}|^2\ket{11}\bra{11},
\nonumber\\
\end{eqnarray}
\begin{figure}[htp]
  \begin{center}
\includegraphics[width=25pc,height=10pc]{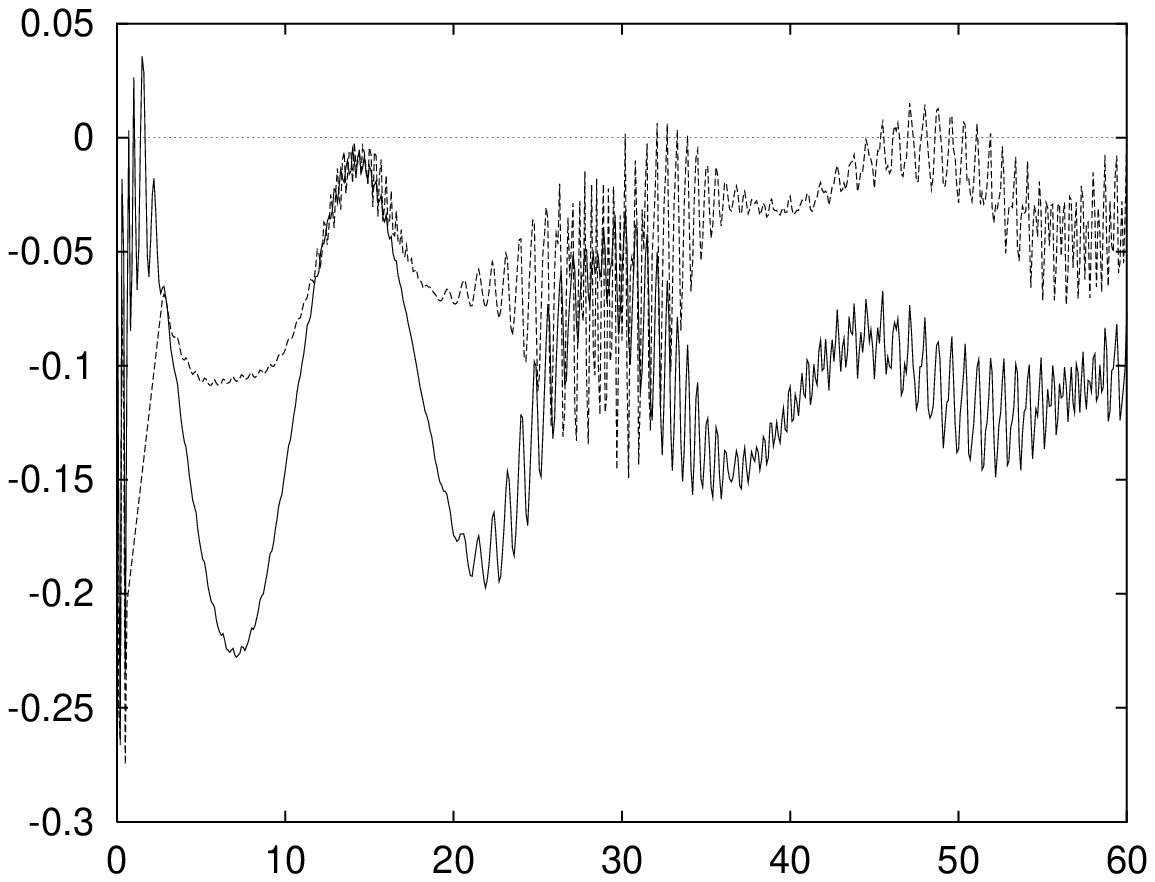}
  \put(-140,-10){$\tau$}
  \put(-30,20){$(a)$}\\
\includegraphics[width=25pc,height=10pc]{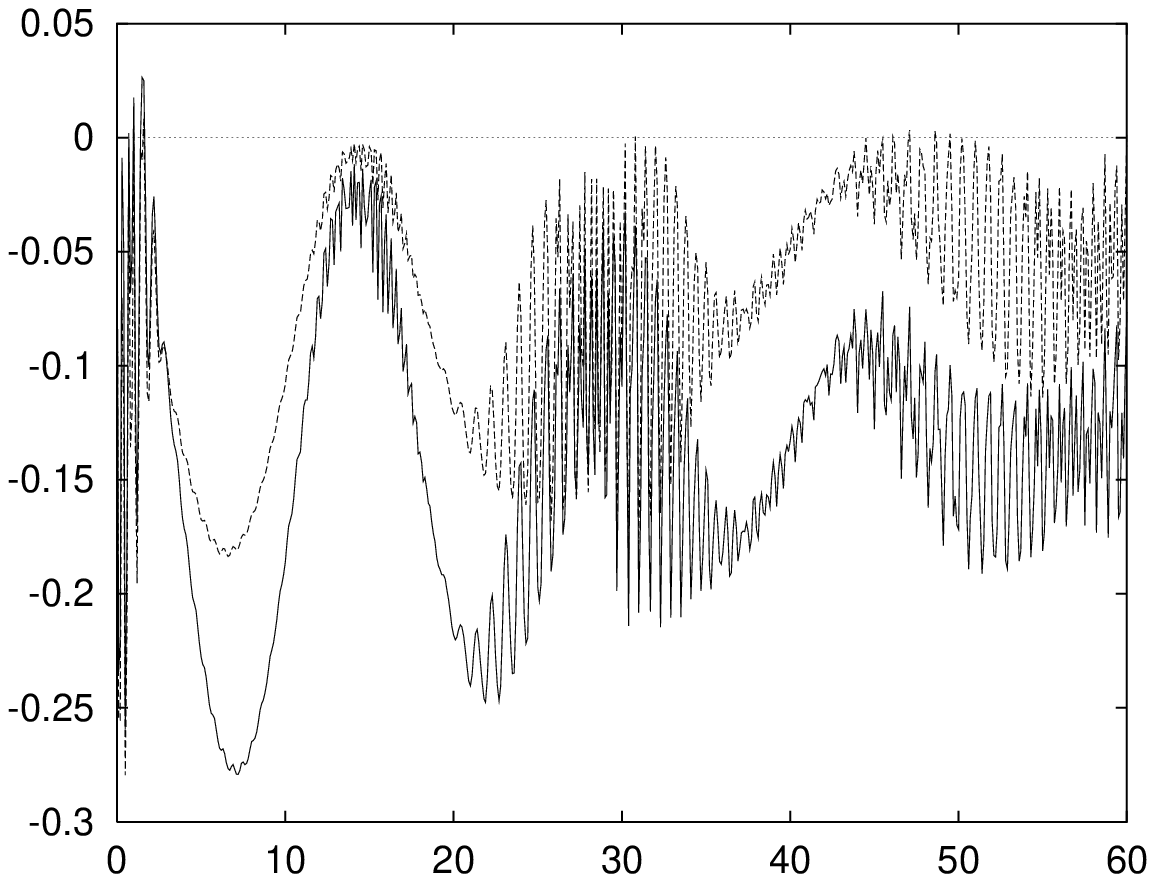}
   \put(-140,-10){$\tau$}
    \put(-30,20){$(b)$}\\
  \caption{The PPT criterion of the quantum channel,
  $\rho_{12}$ with $\bar n = 20$ when Eve sends another
qubit to Bob. the solid curves and the dot curves for the atomic
system initially prepared in excited and superposition product
states respectively. (a) If Eve applies $\sigma_x$ on Alice's
qubit. (b)If Eve applies $\sigma_z$ on Alice's qubit.}
  \end{center}
\end{figure}
Now, let us study the behavior of the channel $\tilde\rho_{12}$
from the separability point of view. For this aim we use the
positive partial transpose criteria (PPT) \cite{peres}. This
criteria states that, a density operator is separable if its
partial transpose is non- negative. If this criteria is violated
then the density operator is entangled. In Fig.$4$, we plot this
criteria for two different values of the initial atomic system. In
Fig.(4a), we plot the PPT criterion when Eve operates by
$\sigma_x$ on Alice's qubit. Starting from an atomic system
prepared initially in a product excited state, one gets an
entangled state once the interaction time goes on. But due to the
instability, this state turns into a separable state in a small
interval of the interaction time, $\tau= [1.4-1.6]$. Then the
state behaves as astable entangled state. On the other hand, if
the atomic system is initially prepared in a superposition product
state, the generated entangled state is instable in along range of
the interaction time, $\tau = [45.4 - 48.9]$. In this interval it
turns into a separable several times.
\begin{figure}
  \begin{center}
\includegraphics[width=25pc,height=10pc]{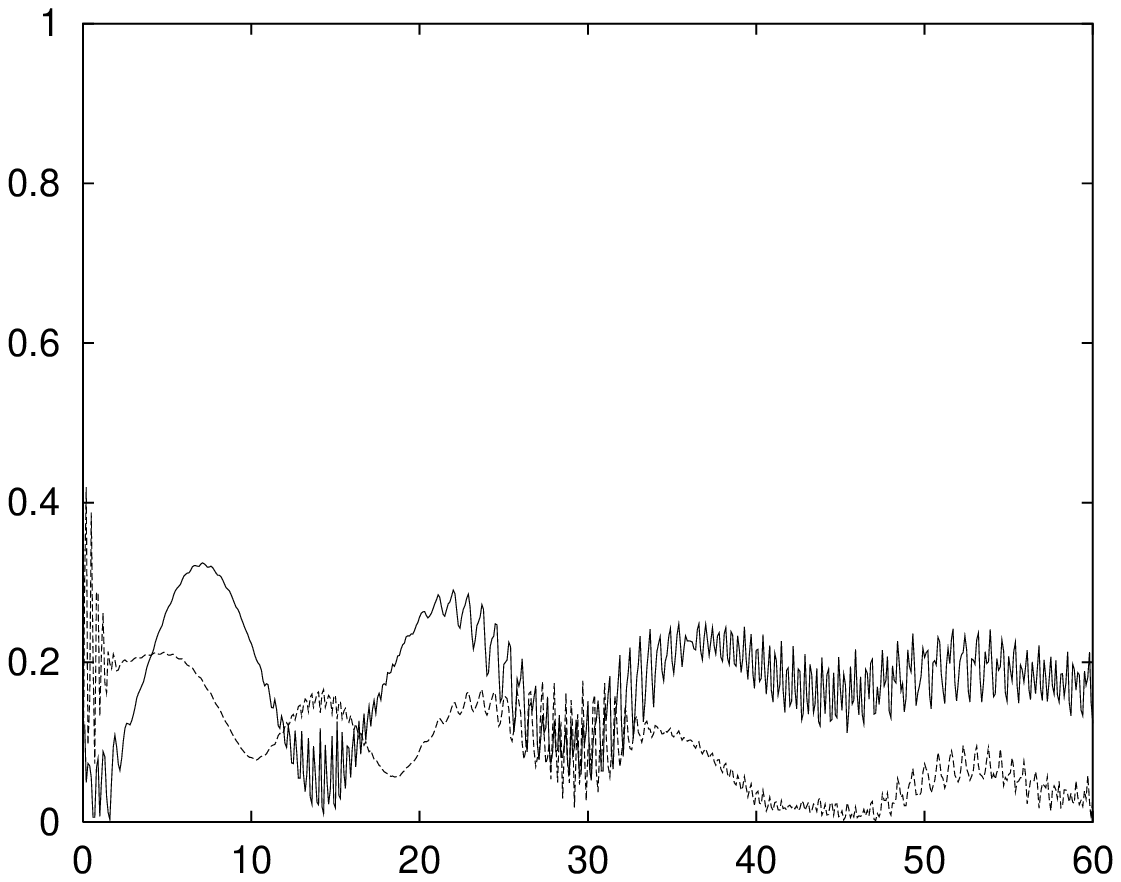}
  \put(-140,-10){$\tau$}
  \put(-30,100){$(a)$}
\put(-280,60){$F$}  \\
\includegraphics[width=25pc,height=10pc]{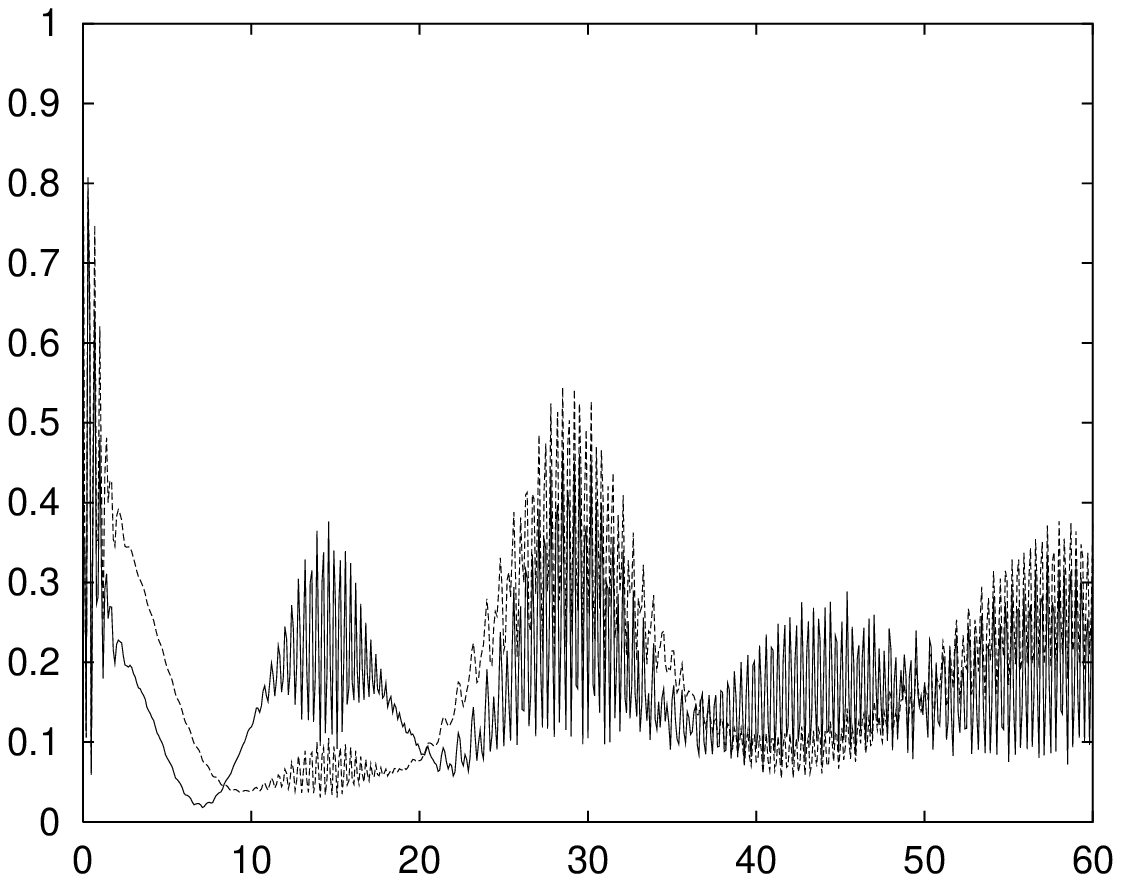}
   \put(-140,-10){$\tau$}
    \put(-30,100){$(b)$}
    \put(-282,60){$F$} \\
  \caption{The fidelity,$F$, of the new channel when Eve send another qubit to Bob. The solid and
the dot curves are for the atomic system is prepared initially in
an excited product state and superposition product states
respectively and $\bar n = 20$. (a) If Alice qubit is rotated by
$\sigma_x$. (b) If Alice qubit is rotated by $\sigma_z$.}
  \end{center}
\end{figure}
In Fig.$(4b)$, we study the separability of a new channel, if Eve
sends another qubit to Bob by rotating by  Alice's qubit by
$\sigma_z$. In this case the generated state is a stable entangled
state expect in a small rang of the interaction time $\tau=
[1.4-1.6]$ where the atomic system is prepared  initially in a
product excited state. But starting by atomic system in a
superposition product states one can generate a more robust
entangled state and never turns into a separable state.

One may ask a question here: how much the new channel related to
the original one?. To answer this question we discuss the {\it
fidelity} of the quantum channel. In this context the fidelity is
defined as \cite{Sch}
\begin{equation}\label{FidT}
F=tr\bigl\{\rho_{12} U_i\otimes
I_2\rho_{12}^{i}U_i^{\dagger}\otimes I_2\bigr\}
\end{equation}
where $i=0,1,2,3$, $U_0=I_1$, is  the unitary operator for Alice
qubit, $I_2$ for Bob qubit, $U_1=\sigma_x, U_2=i\sigma_y$ and
$U_3=\sigma_z$.

In Fig.$(5a)$, we plot the fidelity of the new quantum channel
$\tilde\rho_{12}$, where we consider that Eve applies $\sigma_x$
on Alice's qubit. In this case Bob, will get a state with a small
fidelity. The fidelity decreases for atomic system prepared in a
superposition product states and reaches zero which coincides with
Fig.$(5a)$, where the state is separable. For this strategy one
can expect that, Eve can distill more information from Alice's
massage. In Fig.$(5b)$, the fidelity is plotted for the other
strategy of Eve, i.e when she operates by $\sigma_z$ on Alice's
qubit. In this case the fidelity is better than the previous case
which depicted in Fig.$(5a)$ and never reach to zero. This means
that  in this case the generated states are robust against this
strategy.

\section{Quantum communation}
In this section, we try to employ the generated entangled state to
send a secure information between the partner Alice and Bob.
Assume that Alice has the first atom while the second at Bob's
hand. Let us use a resonator to entangle the two  atoms. For this
purpose the atoms are brought into the cavity for a certain lapse
of time. Then they are removed from the cavity and shared between
(distant) partners (Alice and Bob). The entangled  atom pairs are
distributed and can be used for quantum communication. We use the
dense coding protocol to send the codded information between Alice
and Bob. Also let us assume that the eavesdropper, Eve, would have
access
 the atoms after they have distributed to the partners, and she
 will use the individual attacks strategy. In the following
 subsection we investigate the dense coding protocol. Also we give
 a secure analysis of this communication in Sec.$5.2$.

\subsection{Dense coding Protocol} In this subsection, we use the
generated entangled state to perform the original dense coding
protocol. Since it has been proposed by Bennett and
Wiesner\cite{Bennett}, there are several versions of this protocol
\cite{Bose}. Achieving dense coding protocol via cavity has been
investigated in Ref[12]. Also the dense coding protocol by using
partial entangled state is investigated by Mozes et al
\cite{Mozes}. The most recent practical implementation of dense
coding has been performed by Xing and Gong\cite{Wen}. The original
coding protocol can be described in two steps:

\begin{enumerate}
\item Assume that Alice and Bob share a pair of entangled
particles $\rho_{12}$. Alice can encode two classical bits in her
qubit by using one of the local unitary operations, $U_i$(defined
after Eq.$(12))$ given by . If, we assume that she performs these
operations randomly, then with a probability $p_i$, she codes her
information in the state,

\begin{equation}
\rho_{cod}=\sum_{i=0}^{3}\Bigl\{p_iU_i\otimes
I_2\rho_{12}^{i}U_i^{\dagger}\otimes I_2\Bigr\}.
\end{equation}

\item Alice sends her qubit to Bob, who tries to decodes the
information. To perform this task he makes a joint measurement on
the two qubits, where the two qubits are at his disposal. The
maximum amount of information which Bob can extract from Alice's
massage is bounded by
\begin{equation}
I_{Bob}=\mathcal{S}\Bigl(\sum_{i=0}^{3}{p_i\rho_{12}^{(i)}}\Bigr)-
\sum_{i=0}^{3}{p_i\mathcal{S}\bigl(\rho_{12}^{(i)}\bigr)},
\end{equation}
\end{enumerate}
\begin{figure}[b]
  \begin{center}
\includegraphics[width=25pc,height=10pc]{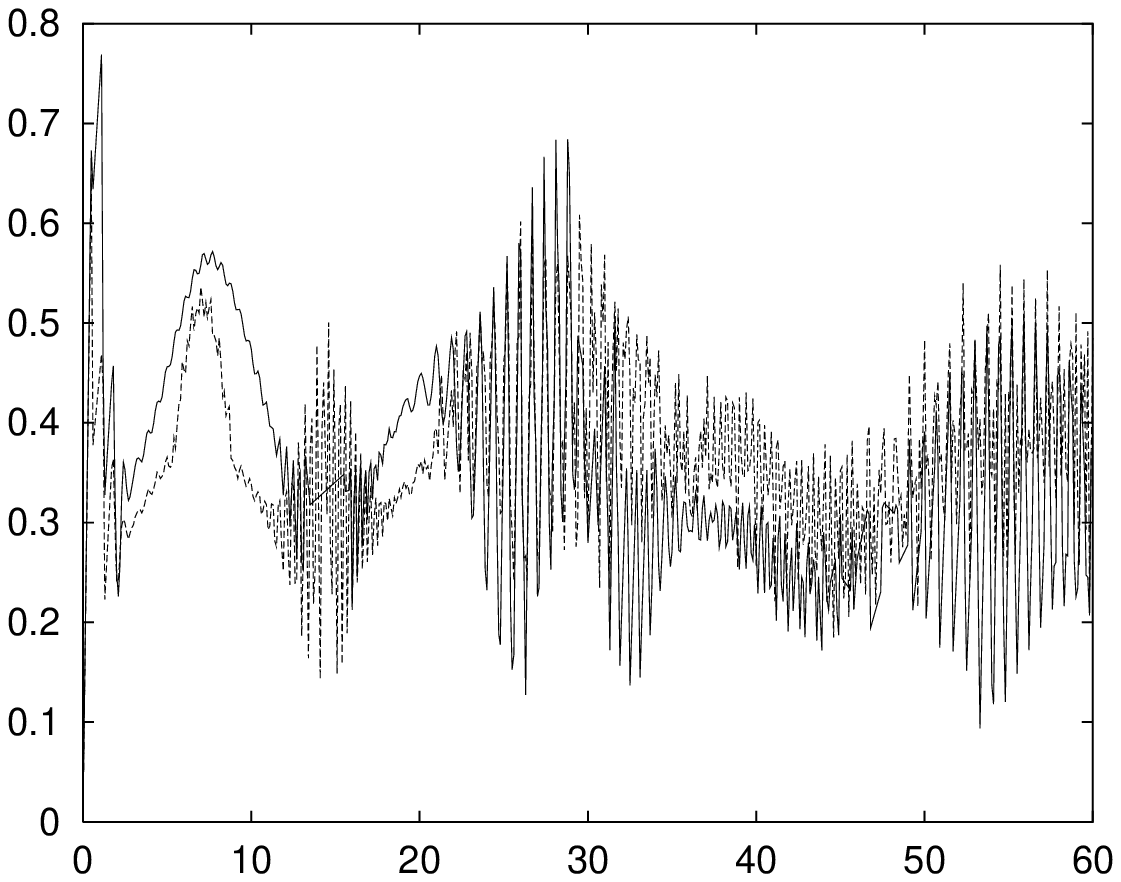}
  \put(-140,-10){$\tau$}
 \put(-300,65){$I_{Bob}$}\\
  \caption{The amount of information decoded by Bob. The dot and the solid curves for the atomic
system is initially prepared in a superposition and excited states
respectively where $\bar n = 20$.}
  \end{center}
\end{figure}
where $\mathcal{S}(.)$, is the Von Numann entropy. In Fig.$6$, we
plot the average amount of information gained by Bob, where we
consider that Alice has used the unitary operator with equal
probability, i.e $p_i = \frac{1}{4}$. From this figure we may
conclude that in some intervals of time Bob can get more
information from the coded massage. On the other hand, if the
Alice and Bob start from atomic system in a superposition state
and use the generated state to code information, it will be better
than if they start from an excited state of the atomic system.

\subsection{Security analysis}
As we have mentation before that Eve will use the individual
attacks strategy. In this case the eavesdropper, Eave can accesses
Alice's atom and resends another one to Bob.  Due to the presences
of the eavesdropper the fidelity, $F$,  of the shared state
between Alice and Bob decreases. In this case the Bob's error
rate, (Disturbance) is defined by $D=1-F$, where
$F=\frac{1}{4}\sum_{i=0}^{3}{tr\bigl\{\rho_{12} U_i\otimes
I_2\rho_{12}^{i}U_i^{\dagger}\otimes I_2\bigr\}}$.
As the
disturbance increases, the fidelity  of the state between Alice
and Bob that govern the probability that they will accept the
transmitted state decreases. On the other hand  Eve's probability
of correctly guessing  more information is increases. In this case
the relevant mutual information between Alice and Eve as a
function of Bob's error is given by \cite{Mohamed},
\begin{equation}
I_{AE}=log_2(2)+(1-D)log_2(1-D)+D log_2{D}.
\end{equation}
The users Alice and Bob can communicate secure and hence they
establish a secret key, if  the Bob's error rate  satisfies the
inequality,
\begin{equation}\label{SecI}
(1-D)log_2(1-D)+D log_2(D)\leq-\frac{1}{2}.
\end{equation}
\begin{figure}[b]
  \begin{center}
  \includegraphics[width=25pc,height=10pc]{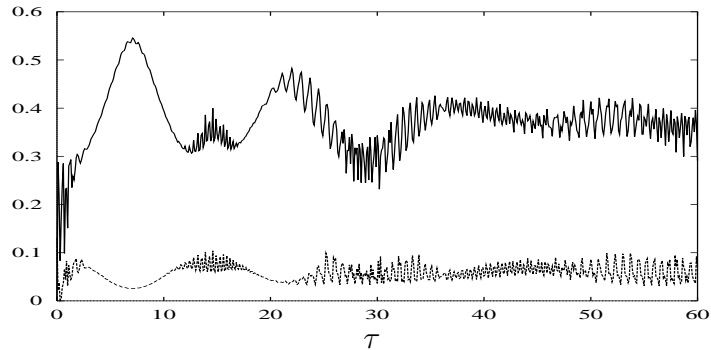}
  \put(-140,-10){$\tau$}
  \caption{The average amount of the mutual information between Alice and Eve $I_{AE}$( dot curves) and
  between Alice and Bob $I_{AB}$(solid curve). The atomic
system is initially prepared in a  excited product states
 where $\bar n = 20$ .}
  \end{center}
\end{figure}

\begin{figure}
  \begin{center}
  \includegraphics[width=25pc,height=10pc]{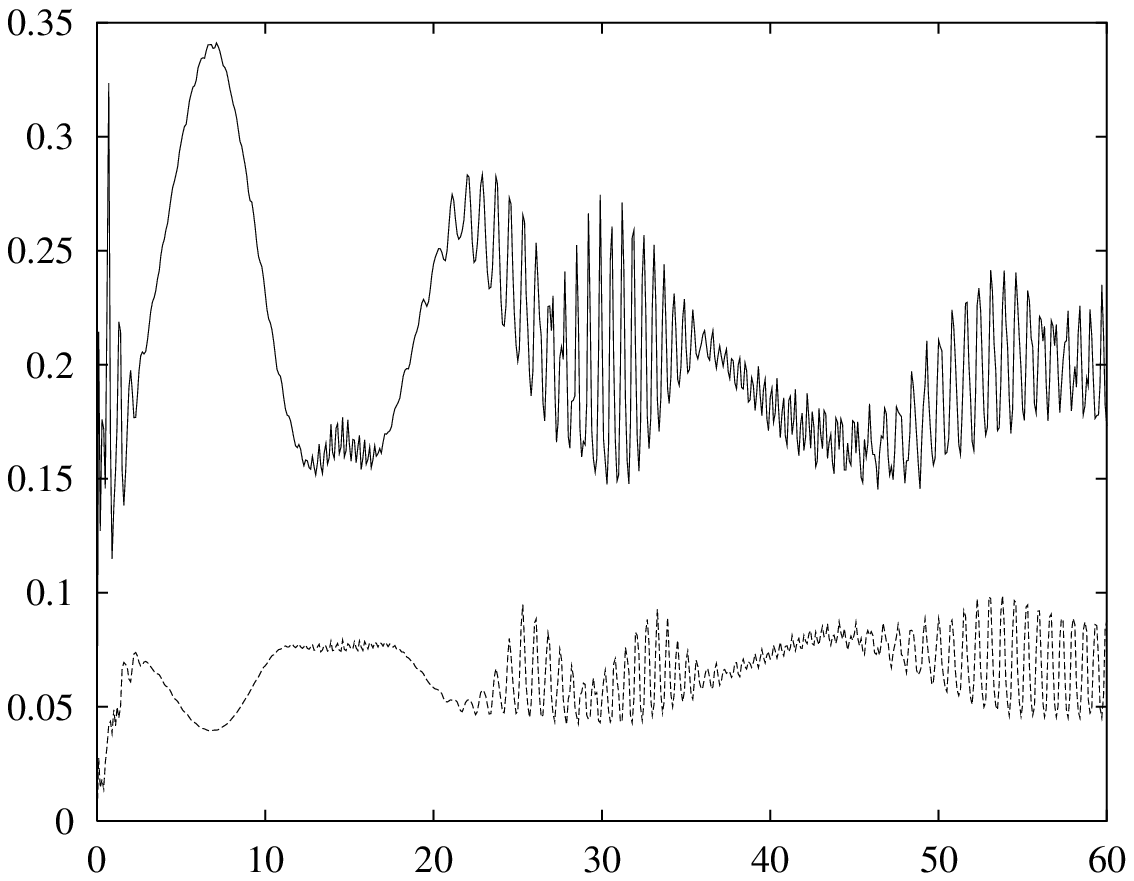}
   \put(-140,-10){$\tau$}
  \caption{The average amount of the mutual information between Alice and Eve $I_{AE}$( dot curves) and
  between Alice and Bob $I_{AB}$(solid curve). The atomic
system is initially prepared in a  superposition  product states
 where $\bar n = 20$ .}
  \end{center}
\end{figure}

\begin{figure}[b]
  \begin{center}
\includegraphics[width=25pc,height=10pc]{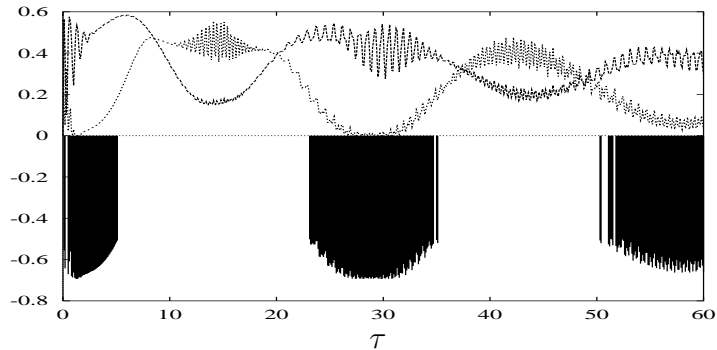}
\put(-140,-10){$\tau$}
   \caption{The mutual information between Alice and Eve $I_{AE}$(dot curve) and between Alice and Bob,
  $I_{AB}$(dash curve). The solid curves represent the secure inequality when it is obeyed.
   Eve applies $\sigma_z$ on the travelling qubit
, $\bar n = 20$ and the  atomic system prepared initially in a
superposition  state.}
  \end{center}
\end{figure}

This inequality gives bounds on Bob's permissible error rate in
the case of individual attacks.

Fig.$7$ shows the behavior of the  average amount of the  mutual
information between Alice, Eve ($I_{AE})$ and Alice, Bob
($I_{AB}$), where we assume that the initial atomic system is
prepared in a excited product case. This information is plotted
when the inequality of the security, (\ref{SecI}) is obeyed. It is
clear that in some intervals of time, the mutual information
between Alice and Eve, $I_{AE}$ decreases while the mutual
information between Alice and Bob, $I_{AB}$ increases. Although in
some intervals of time $I_{AE}$  increases in the expanse of
$I_{AB}$, Alice and Bob  still communicate in a secure way.
 Fig.$8$, shows the behavior of  $I_{AB}$ and $I_{AE}$, where the
atomic system is prepared in a superposition product state and the
secure inequality (\ref{SecI}) is satisfied. By comparing
Fig.$(7)$ and Fig.$(8)$, we can see that $I_{AB}$, which is
depicted in Fig.$(7)$ is much larger than that in Fig.$(8)$. So,
starting from atomic system prepared in excited product state, the
partner can communicate safely and the secure information is much
larger.

In Fig.$(9)$, a special case is considered where we assume that
Eve, causes  a phase error on the travelling qubit. In this
figure, the behavior of the mutual information $I_{AB}$, $I_{AE}$
and the inequality of security (\ref{SecI}) are plotted. It is
clear that, for some intervals of time $I_{AE}<I_{AB}$ and the
inequality of security is obeyed. So in these intervals Alice and
Bob can communicate in a secure way. Although  $I_{AE}<I_{AB}$ in
some  intervals of time, as an examples $[5.2-8.2]$ and
$[20.2-23.]$, but the channel is insecure. In this case Alice and
Bob will not accept the probability of the transmitted
information. For some other intervals of time $I_{AE}>I_{AB}$ and
the inequality of security is violated. This means that through
these intervals the channel is insecure  and the partner, Alice
and Bob can not communicate safely.

\section{conclusions}

In this paper, we have considered the system of two two-level
atoms interacting with a cavity field. It is shown that,
generating entangled states with high degree of purity and hence
high degree of entanglement depends on the initial state of the
atomic system. For our system this is achieved with the
superposition state. The dynamics of the purity of the individual
atoms show the swapping phenomena. The purity of  one of  qubit
can be purified at the expense of the other qubit through the
dynamics of the purity swapping. Also, as one increases the values
of the mean photon number, the degree of purity increases and
consequently the degree of entanglement.

 The robustness and the fragile of the channel are
investigated in the presences of the individual attacks. We find
that for some strategy of Eve the channel is fragile. In  this
case Eve can distill some information from the coded massage. On
other strategy, the channel is more robust and the eavesdropper
can not get more information. Also  if we start with atomic system
prepared  initially in excited state, one can generate entangled a
more robust entangled states.

 Finally, we
employ the generated entangled state to perform the original dense
coding protocol. It is possible to send a codded massage from
Alice to Bob with reasonable fidelity. This fidelity depends on
the structure of the initial atomic system.
 It has been shown that choosing the atomic system initially prepared in a
superposition state is much better.

 We show that the  average amount of the coded information can be transmitted
between the users securely, where the  inequality of security is
obeyed. It is clear that, although the average codded information
is better if the partner start with a superposition state, the
possibility of the secure communication is decreases.
 Also, an example is given, where we assume that Eve applies the shift error operator.
 We determine the intervals of time in which the channel is
secure and the partner can use it safely, where the inequality of
security is tested.

\section*{Acknowledgments}
I would like to thank the referees for subjective reports and
helpful comments. Also, many thanks to Prof. S. Hassan and Dr. M.
abdel-Aty for their useful discussions.

\end{document}